\documentclass{article}
\usepackage{amsmath}
\begin{document}

\centerline{\Large\bfseries Radiometric force in dusty
plasmas}\bigskip
\centerline{\normalsize A.M.Ignatov, Sh.G.Amiranashvili}\bigskip
\centerline{\normalsize\em General Physics Institute, Moscow 117
942}\medskip

\begin{abstract}
A radiofrequency glow discharge plasma, which is polluted with a
certain number of dusty grains, is studied. In addition to various
dusty plasma phenomena, several specific colloidal effects should
be considered. We focus on radiometric forces, which are caused by
inhomogeneous temperature distribution. Aside from thermophoresis,
the role of temperature distribution in dusty plasmas is an open
question. It is shown that inhomogeneous heating of the grain by
ion flows results in a new photophoresis like force, which is
specific for dusty discharges. This radiometric force can be
observable under conditions of recent microgravity experiments.
\end{abstract}

PACS number(s): 52.25.Vy, 52.25.Ya, 52.90.+z, 82.70.Dd
\medskip

Numerous industrial applications, such as material proceeding, has
triggered active research on the phenomena associated with dust
dynamics in a low-pressure glow discharge. A dusty plasma
\cite{Fortov, Tsytovich} is formed by introducing micron-sized
grains in a plasma. The grains are negatively charged due to the
higher mobility of electrons with respect to ions. Then the medium
is composed of particles with fixed charge (electrons and ions)
and variable charge (dust grains). Typically, the microspheres can
be easily charged to $10^4-10^5$ electron charges. In a
low-temperature radiofrequency discharge the grains usually remain
electrically suspended in the sheath above the electrodes
\cite{ChuLin}. Here the gravity is exactly balanced by the
electric force. The grains form ordered lattice structures, known
as Coulomb crystals.

The plasma boundary near the electrodes is characterized by highly
non-equilibrium conditions. In particular, the grains undergo a
supersonic (Bohm criterion) ion flow, resulting both in an ion
drag force and  specific attractive forces between the
grains~\cite{Vladimirov}. The neutral gas causes an important
additional class of forces, which are related to the density and
temperature gradients. Aside from thermophoresis \cite{Havnes,
Perrin} the role of such forces is an open question. In this rapid
communication we focus on a radiometric forces. We found that ion
flows result in inhomogeneous temperature distribution of the
grain surface. Then the interaction with the neutral gas results
in a force similar to photophoresis \cite{Preining}, but it is
provided by a plasma recombination at the grain surface.

Let $Q$ denote the energy released in each act of ion
recombination. Typically, the value of $Q$ is of the order of ten
eV. We assume that this energy is absorbed by the grain and
results in the heating of its surface. Recall that the ion flow in
the sheath near the electrodes is strongly non-isotropic. The
widely spread approximation is that zero-temperature ions move
towards the electrode with the same supersonic velocity $Mc_s$,
where $M>1$ is Mach number, and $c_s$ is the ion sound velocity.
Then the energy flux per a unit square of the grain surface
oriented normally to ion velocity, can be estimated as
\begin{equation}\label{flux}
J_0=\alpha n_iMc_sQ,
\end{equation}
where $n_i$ is the ion number density and dimensionless factor
$\alpha$ takes into account the attraction of ions with the
negatively charged microsphere. This will increase the resulting
ion flux. Typically $\alpha\approx2$.

The steady energy flux affects only one side of the grain
resulting in some inhomogeneous heating. The stationary
temperature distribution inside the grain can be evaluated by
means of the the Fourier's equation
\begin{equation}\label{Fourier}
\mathop{\mathrm{div}}\biggl[\kappa\mathop{\mathrm{grad}}T(\mathbf
r)\biggr]=0,
\end{equation}
where $\kappa$ is the grain thermal conductivity.
Eq.(\ref{Fourier}) should be supplemented with a boundary
condition. We characterize the grain surface by a unit vector
$\mathbf n$ directed outwards. The boundary condition reads
\begin{equation}\label{boundary}
J_s=\kappa\left(\frac{\partial T}{\partial \mathbf
n}\right)_s+\sigma(T_s-T_0), \end{equation}
where the subscribe $s$ recalls that this equation is applied to
the grain's surface only. The function $J_s$ is an external energy
flow onto the grain, which is caused by the recombination. The
first term in the right-hand side of Eq.(\ref{boundary})
represents the energy flow in the interior of the grain. We
emphasize that the the grain permanently interacts with the
neutral gas, which is assumed to be uniform, with some temperature
$T_0$ differing from the grain's surface temperature $T_s$. The
interaction results in heat transfer, which is described by the
second term in the right-hand side of Eq.(\ref{boundary}). The
rate of this natural cooling is characterized by the factor
$\sigma$.

For a microsphere suspended above the electrode the external
energy flow is taken in the form
\begin{equation}\label{flow}
J_s=\left\{\begin{array}{cl}-\mathbf{J_0}\mathbf{n}&\mbox{for top
half,}\\ 0&\mbox{for bottom half.}\end{array}\right.
\end{equation}

Eqs. (\ref{Fourier}-\ref{flow}) provide a complete description of
the temperature distribution inside the gain. For a uniform
spherical grain they are readily solvable in terms of Legendre
polynomials.

To proceed it is necessary to specify the collision of the neutral
molecule with a grain surface. We undertake the assumption of
complete energy accomodation. Typically the size of a grain is
small, as compared to the mean free path of the neutral molecule.
Then the particle distribution in the vicinity of the grain can be
taken as a combination of two Maxwell functions:
\begin{equation}\label{distribution}
f(\mathbf r,\mathbf v)= \left\{\begin{array}{cl}f_M(T_0)&\mbox{for
$\mathbf{nv}<0$,}\\ f_M(T_s)&\mbox{for $\mathbf{nv}>0$.}
\end{array}\right. \end{equation}

In should be noted that the number density of the neutral
particles moving towards and outwards the grain ($n_0$ and $n_s$
respectfully) is different $n_0\sqrt T_0=n_s\sqrt T_s$ in
according to the conservation of the particle flux. Using
Eq.(\ref{distribution}) one obtains
$$\sigma=n_0\sqrt{\frac{2T_0}{\pi m}},$$
as well as the neutral gas pressure
$$P=\frac12n_0\sqrt{T_0}(\sqrt{T_0}+\sqrt{T_s}),$$
where $n_0$ is identical to the number density of the neutral gas.

Note, that the pressure depends on the local surface temperature.
Due to the inhomogeneous temperature distribution the interaction
with the neutral gas results in a force expressed as the integral
$$ \mathbf F = \iint(-P\mathbf n)\,dS$$
over the grain surface.

In a case of a uniform spherical particle the integration can be
carried out analytically. The cumbersome final expression
simplifies greatly in the ultimate case of $T_s-T_0\ll T_0$, which
is the only of interest here. The resulting force takes the form
\begin{equation}\label{final}
F=\frac{\pi
R^2n_0J_0}{6\left(\sigma+{\displaystyle\frac{\mathstrut\kappa}R}\right)},
\end{equation}
where $R$ is a radius of a microsphere.

Eq.(\ref{final}) closely resembles the known relation for the
photophoresis force \cite{Preining, Hidy}. The difference is due
the different physical meaning of the term $J_0$. In the case of
the photophoresis the inhomogeneous heating is a result of
external radiation. In a dusty plasma the force results from the
ion recombination.

In some experiments \cite{France} the dusty specie is formed by
relatively big coreless grains. Let the inner radius of that
hollow microsphere be $\epsilon R$ with constant $\epsilon<1$.
Assuming that the energy flux through the inner surface is
negligibly small one can easily obtain the following more general
expression
\begin{equation}\label{efinal}
F=\frac16\frac{\pi R^2n_0J_0}{\sigma(1+\frac12\epsilon^3)+
{\displaystyle\frac{\mathstrut\kappa}R(1-\epsilon^3)}}.
\end{equation}

For typical conditions of dust dynamics the first term in the
denominator of Eq.(\ref{efinal}) is small as compared to the
second term. Then
\begin{equation}\label{reduced}
F=\frac{\pi R^3n_0J_0}{6\mathstrut\kappa(1-\epsilon^3)}.
\end{equation}

The ratio of this radiometric force to gravity is independent of
the radius of the grain. Generally, it is extremely small. Let us
compare the radiometric force with the ion-drag force, which is
believed to play an important role under conditions of recent
microgravity experiments with dusty plasmas \cite{Morfill}.

The ion-drag force is estimated as $F_{\mathrm{drag}}=\alpha\pi
R^2 n_im_i(Mc_s)^2$, where the physical meaning of factor $\alpha$
is identical with that in Eq.(\ref{flux}). Hence
$$\frac{F}{F_{\mathrm{drag}}}=\frac{Rn_0Q}{6\kappa
m_iMc_s(1-\epsilon^3)}.$$

Typically $M\approx1$, electron temperature $T_e=2.5$~eV,
$Q=15.8$~eV (Argon), and $\kappa\approx10^{-2}$~Wt/cm$\cdot$grad
(glasses). We take $n_0=10^{16}$~cm$^{-3}$, and
$1-\epsilon^3=1/30$. Then $F/F_{\mathrm drag}=1$ if
$R=10^{-2}$~cm. The effect therefore should be observable for
relatively large hollow dusty grains under microgravity
conditions.

\section*{\normalsize ACKNOWLEDGMENTS}

This work was performed under the financial support granted by the
NWO foundation. We also acknowledge the support from Integration
foundation, project \#A0029.

\end{document}